\documentclass[12pt,letter]{article}
\usepackage{epsfig}
\usepackage{amssymb}

\begin{document}

\begin{center}
{\Large \sl Search for QCD-instanton-induced effects in deep inelastic
electron proton scattering at HERA$^\ast$}
\end{center}

\begin{center}
Gerd W. Buschhorn\\
Max-Planck-Institut f\"ur Physik\\
(Werner-Heisenberg-Institut)\\
80805 M\"unchen, Germany
\end{center}

\section{Introduction}

QCD, in contrast to QED, is not exhausted by perturbation theory. While QCD
perturbation theory appears to be very successful at higher
momentum transfers where the strong coupling constant becomes small, some
of the most fundamental properties of hadronic physics like color
confinement and hadronization of quarks and gluons, characteristic for the
low momentum transfer region, are not accessible to perturbative methods.

QCD is a nonabelian gauge theory. In 1975, Polyakov et al. \cite{1} made the
important discovery that such theories in their pure gauge form,
i.e. without the involvement of spontaneous symmetry breaking, have
localized classical solutions in Euclidean space-time. Since the objects
described by these solutions are localized in 3+1 dimensions, they describe
``events'' rather than normal particles which are localized in 3
dimensions. Accordingly, they were named pseudoparticles by Polyakov
[1] and instantons by 't Hooft \cite{2}. The replacement of time $t$ by
$ix_4\hspace{0.3cm}(\rm{real}\hspace{0.3cm}x_4)$ by going into Euclidean
space corresponds to the method of calculating tunneling amplitudes in the
nonperturbative WKB approximation, where the characteristic exponential
suppression factor $\exp(-8\pi^2/g^2)$ is obtained from solving the
classical equations for imaginary time. Instantons are, therefore,
interpreted as tunneling events. The tunneling, of course, takes place in
real (Minkowskian) space, although it is actually computed in Euclidean
space. As was first realized by Ringwald and Espinosa \cite{3}, the
exponential suppression of tunneling processes may be overcome at high
energies by multiple emission of gauge bosons accompanying fermion
production. 

The tunneling events violate a conservation law that is obeyed by
perturbative solutions: For the classical (BPST) solutions there exists a
charge like quantity

\begin{equation}
\int d^4x F_{\mu \nu} \tilde{F}_{\mu \nu} = \pm \frac{32\pi^2}{g^2}
\end{equation}

where $F_{\mu\nu}$ is the gauge field strength tensor and
$\tilde{F}_{\mu\nu}$ its dual tensor. Since from the Adler-Bell-Jackiw
axial triangle anomaly \cite{4} $(g^2/16\pi^2)F_{\mu\nu}\tilde{F}_{\mu\nu}$
equals the anomalous divergence of the axial-vector divergence, tunneling
events i.e. instantons connect states of different axial charge $Q_5(t) =
\int d^3x J_0^A(x,t)$ and one obtains

\begin{equation}
\Delta Q_5 = Q_5(\infty) - Q_5(-\infty) = \int_{- \infty}^{+ \infty}dt \int
d^3x \partial_0 J_0^A(x) = \pm 2 .
\end{equation}

This relation holds separately for each flavour and for $N_f$ flavours one
has

\begin{equation}
\Delta Q_5 = \pm 2N_f .
\end{equation}

Due to their topolocical nature (1) and (2) hold in Euclidean as well as in
Minkowskian space-time. The violation of the conservation of axial charge
i.e. helicity is a most significant property of instanton events.

The relevance of instantons for different aspects of hadronic physics has
been reviewed recently \cite{5}. In the following, we discuss
recent work on the role of QCD-instantons in deep inelastic electron proton
scattering. 

\section{Instantons in Deep Inelastic Electron Proton Scattering}

Balitsky and Braun \cite{6} first showed that the contribution of instanton
($I$)-induced processes to deep inelastic electron proton scattering from a
real gluon calculated in Euclidean space and continued to Minkowski space, 
rises very rapidly with decreasing Bjorken-$x$.  Their calculations were
restricted, however, to $x > 0.3 - 0.35$. A detailed theoretical 
\cite{7,8,9} and phenomenological \cite{10,11} investigation of deep
inelastic scattering at HERA has been pursued by Ringwald, Schrempp and
collaborators.

The leading $I$-induced process contibuting to deep-inelastic
$ep$-scattering, given by a hard quark originating from the virtual
photon and reacting in the presence of an instanton with a gluon from the
proton, is shown in fig. 1.


\begin{figure}[htbp]
 \begin{minipage}[t]{6.5cm}
     \makebox[0cm][t]{}        
     \begin{center}
       \includegraphics[width=0.95\textwidth]{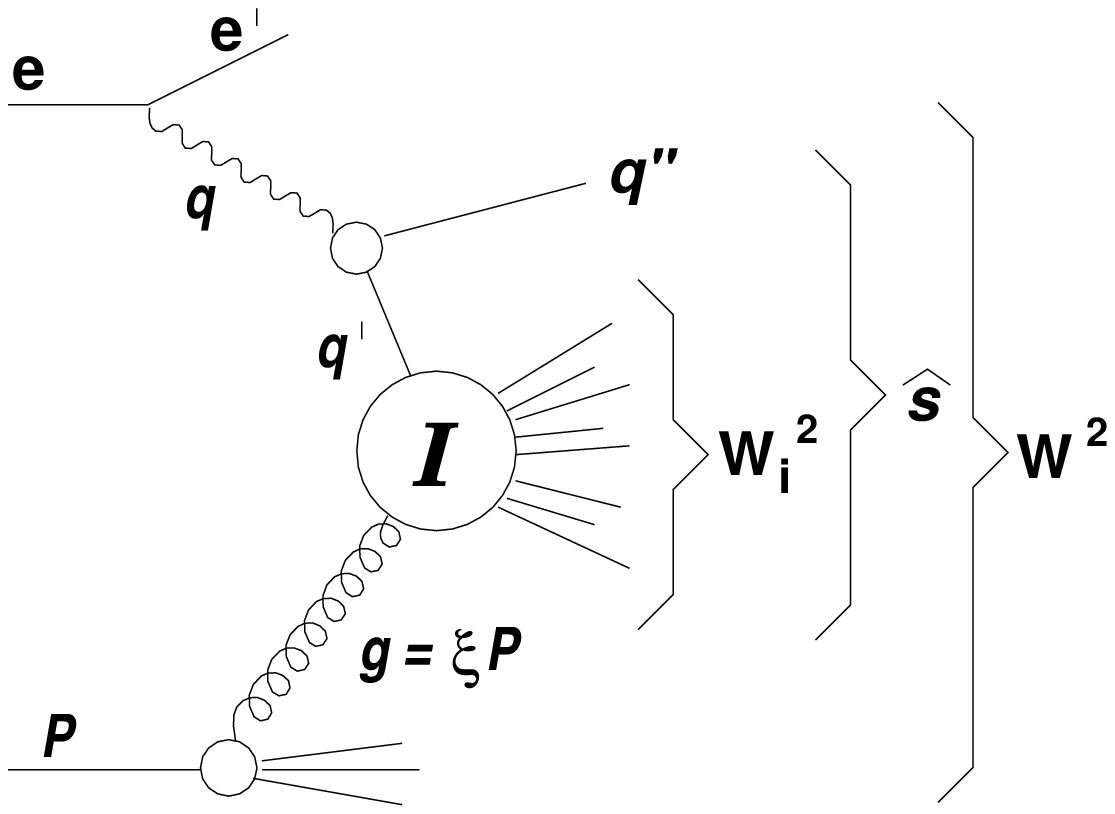}
     \end{center}
 \end{minipage}
 \hspace{0.5cm}
 \hfill
 \begin{minipage}[t]{6.5cm}
   \vspace{0.1cm}
   \label{abb01}
   \caption[]{\small Leading graph for instanton-induced contribution to deep
              inelastic electron proton scattering. It denotes the
              instanton subprocess with the observables $Q'^2 = -q'^2 =
              -(q-q')^2$, $x'=Q'^2/2(g\cdot q')$, $W_i^2 = (q'+g)^2=Q'^2
              (1-x')/x'$.}
  \end{minipage}
\end{figure}


The corresponding contribution with a quark originating from the proton is
suppressed by $\alpha_s^2$ and can be neglected in the kinematical region
of interest (see below), where the gluon density is high. 

The inclusive deep-inelastic $ep$ cross section in $I$-perturbation theory
in the Bjorken limit can be approximated as

\begin{equation}
\frac{d\sigma_{ep}}{dx'dQ'^2} =
\frac{d{\cal{L}}_{qg}}{dx'dQ'^2}\sigma^I_{gg}(x', Q'^2)
\end{equation}

where $d {\cal{L}}_{qg}/ d x'd Q'^2$ is a differential luminosity
corresponding to a con\-vo\-lu\-tion-like integral over photon flux, quark flux
and gluon density and $\sigma^I_{qg}$ is the total cross section of the
$I$-induced quark-gluon subprocess.

The critical part of (4) is the calculation of $\sigma_{qg}^I$ which
contains all $I$-dynamics. Outlining the essential dependencies, it can be
written as \cite{8,9}
\begin{equation}
\sigma_{qg}^I = \int^\infty_0 d\rho D(\rho) \int^\infty_0
d\bar{\rho}D(\bar{\rho}) \int
d^4Re^{-Q'(\rho+\bar{\rho})} e^{i(p+q')\cdot R} \int dU e^{-
\frac{4\pi}{\alpha_s}\Omega
\left(\frac{R^2}{\rho\bar{\rho}},\frac{\bar{\rho}}{\rho}, U\right)}\{...\}
\end{equation}

where the integration is performed over the ``collective coordinates'' of
the instanton i.e. the size $\rho$ weighted with the density distribution
function $D(\rho)$, the $I-\bar{I}$-distance vector $R$ and the $I$-
$\bar{I}$-relative colour orientation $U$. $I$ as well as
$\bar{I}$ enter here since the cross section results from the modulus
squared of the amplitude for processes induced by single $I$
resp. $\bar{I}$. The function $\Omega$ with the large factor $-4\pi/\alpha_s$
contains multiple emission of gluons. $D(\rho)$ is known in
$I$-perturbation theory for $\alpha (\mu_r)\ln (\rho \mu_r) \ll 1$ where
$\mu_r$ is the renormalization scale. $D(\rho)$ has a powerlaw behaviour.
\begin{equation} 
D(\rho) \sim \rho^{6 - \frac{2}{3} n_f + 0(\alpha)}
\end{equation}
 
which in general results in a breakdown of $I$-perturbation theory for
large values of $\rho$. In deep inelastic scattering, however, the exponential
term [7] containing $Q'$ supresses the contribution form large size
instantons and is the key for the applicability of $I$-perturbation theory.

For calculating the cross section for deep inelastic scattering DIS at HERA,
the range of collective coordinates has to be known in which
$I$-perturbation theory can be relied on. This range of
collective coordinates then has to be translated into the relevant
DIS-variables.

For estimating the range of validity of $I$-perturbation theory QCD lattice
results from the UKQCD collaboration [12] have been extremely 
useful. Fig.~2a shows a comparison of $I$-perturbation theory with
lattice-QCD in the continuum limit [9] for the $I$-size distribution. The
normalization is practically independent on the renormalization scale
$\mu_r$, but strongly dependent on the value of $\Lambda_{\overline{MS}}$,
for which the recent accurate result from the 
ALPHA collaboration \cite{13} $\Lambda_{\overline{MS}(n_f=0)}=(238\pm19)
\textrm{MeV}$ has been taken. It has to be noted that the
pertubative result is essentially parameter free. Fig.~2b shows the
corresponding comparison for the $I-\bar{I}$-distance
contribution. Very good agreement down to
$I-\bar{I}$-distances $R/<\rho^-> \simeq 1$ is observed.

In the deep inelastic
regime the integrals over collective coordinates are dominated by a unique
saddle point \cite{8} which relates the collective coordinates to the Bjorken
variables $Q', x'$ of the $I$-subprocess as
\begin{equation}
\rho^*, \bar{\rho}^* \sim 1/Q', \hspace{0.5cm} R^{*2} \sim 1/(p+q')^2
\rightarrow R^{*2}/\rho^*\bar{\rho}^* \sim Q'^2 / (p+q')^2 = x'/1-x'.
\end{equation}

The limits on the collective observables then transform into limits of the
DIS variables in the following way:
\begin{equation}
\begin{array}{lcl}
\rho^* \le 0.30 - 0.35 fm && \hspace{0.4cm}
Q'/\Lambda_{\overline{MS}} \ge 30.8\\ 
&\rightarrow&  \\
R^*/\rho^* \ge 1 &&\hspace{0.4cm} x' \ge 0.35
\end{array}
\end{equation}

\vspace{0.3cm}

\begin{figure}[h]
\begin{center}
\epsfig{file=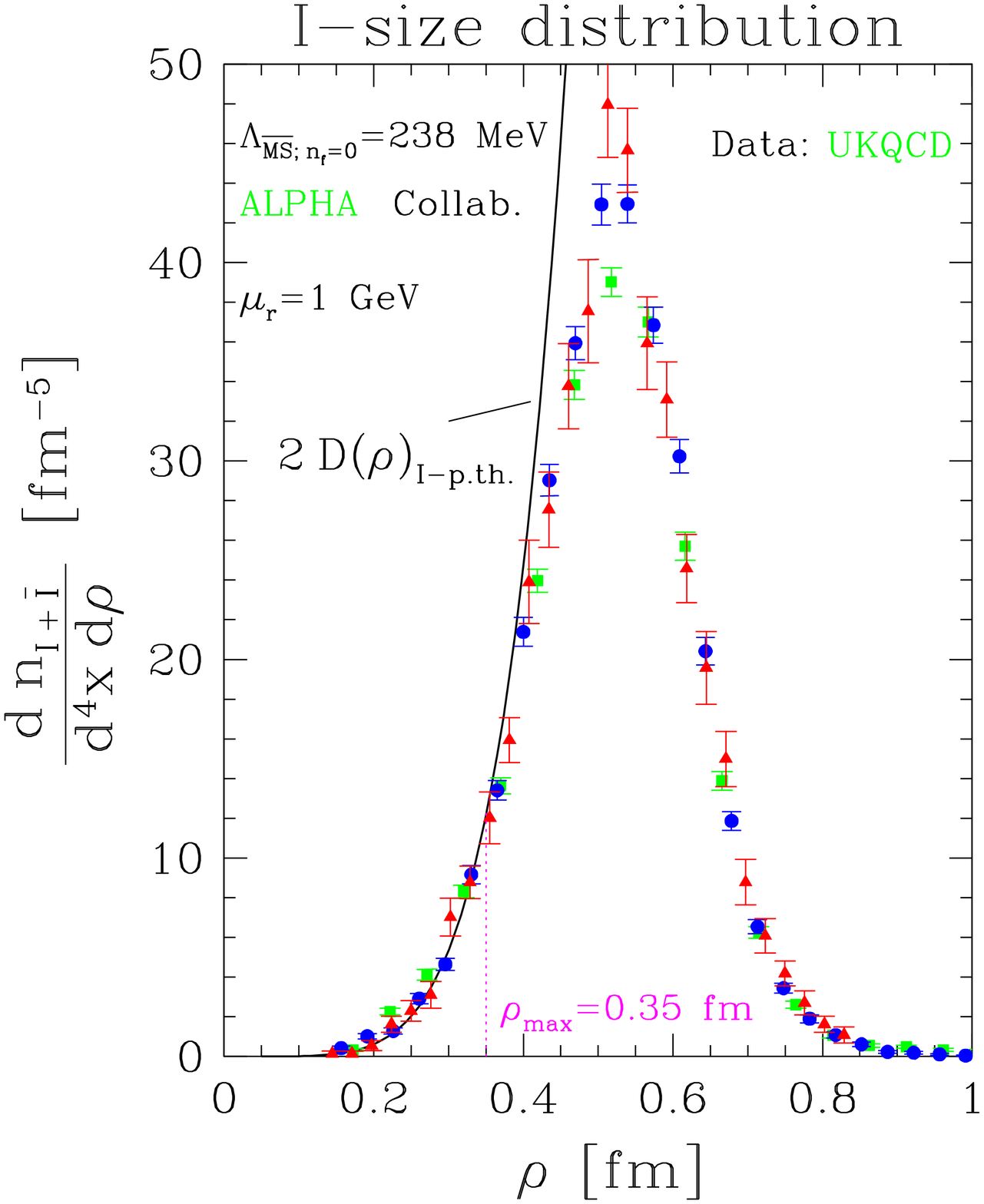,width=6.5cm}
\epsfig{file=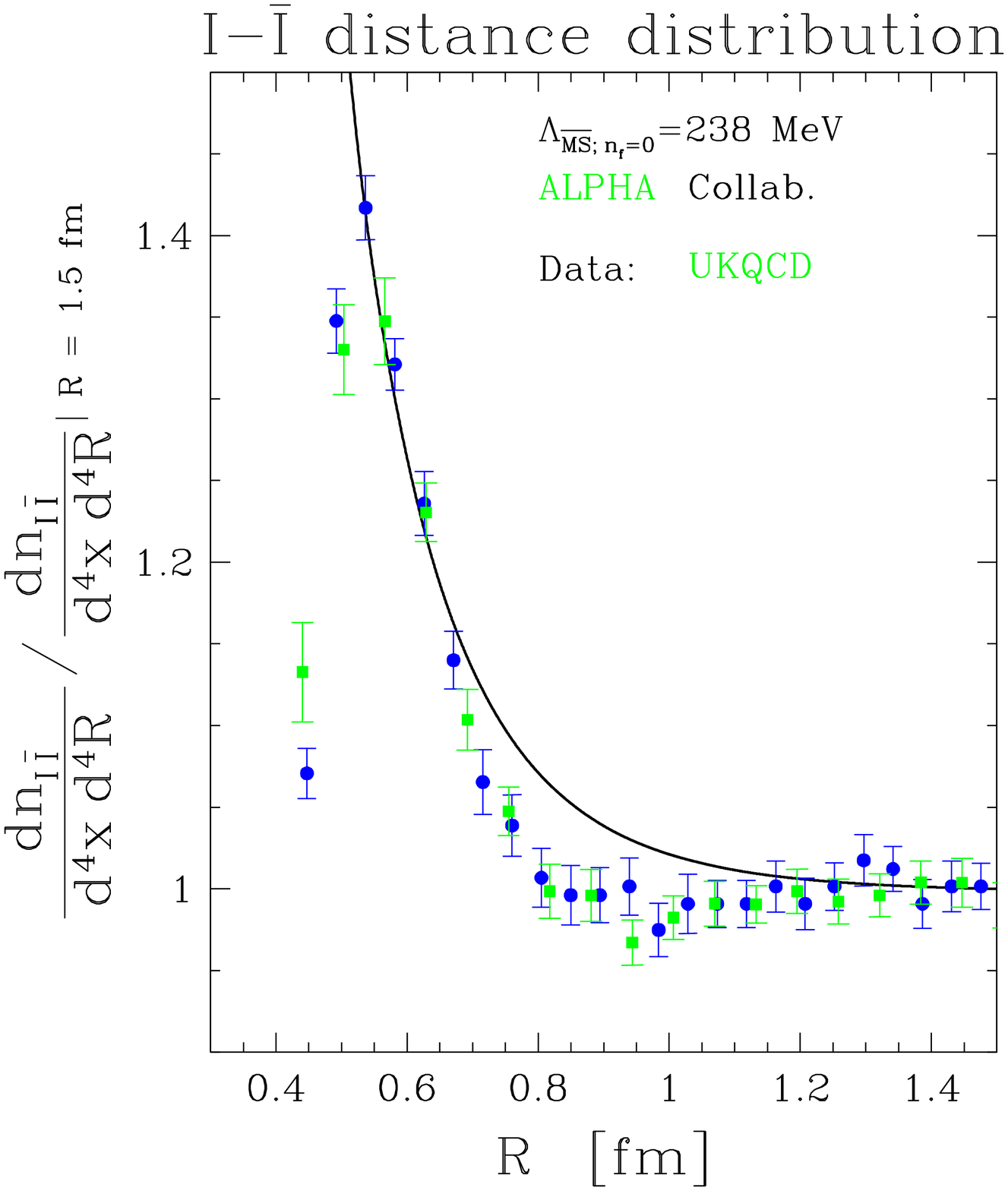,width=6.5cm}
\end{center}
\label{abb02}
\caption{\small Comparison of ``collective coordinates'' obtained from
instanton-perturbation theory (curve) with the continuum limit of
equivalent lattice QCD data (points): a) $I(\bar{I})$-size distribution; b)
$I-\bar{I}$-distance distribution, normalized to the value at $R=1.5
fm$. From Refs. \cite{9,19}}
\end{figure} 

With these cuts, additional general cuts of $x > 10^{-3}$ and $0.1 < y <
0.9$ and a ``technical cut'' of $Q'^2_{min} = Q^2_{min}$, in order to
suppress contributions from nonplanar graphs which are hard to calculate, a
total cross section for $I$-induced events of $29 \pm 10$ pb is
obtained, where the errors refer to the uncertainty in
$\Lambda_{\overline{MS}}$ only. For a total integrated luminosity of about 100
pb$^{-1}$, which will be reached for each of the HERA experiments H1 and
ZEUS by fall 2000, this corresponds to about 3.000 $I$-events. Since the
total cross section for standard DIS events is several orders of magnitude
higher it is evident, however, that a search of $I$-induced effects has to
make use of specific event characteristics.

\section{Search Strategies for $I$-induced Events at HERA}

The Monte Carlo generator QCDINS \cite{14}, developed on the basis of the
perturbative model discussed above for the detailed simulation of
$I$-induced events, yields the following event characteristics \cite{10}:  

\begin{itemize}
\item a jet (``current jet'') from the outgoing quark in the primary photon
interaction (compare fig.~1)
\item after removal of the current jet a uniformly populated band of
hadrons in $\eta - \phi$-space (with the pseudorapidity $\eta$ and the
azimuthal angle $\phi$) resulting nearly in isotropic decay in the hadronic
CMS (defined by ${\bf{q}}+{\bf{P}}=0$)
\item high parton multiplicity (for 3 flavours: 3 $(q\bar{q})$-pairs; at
$\alpha_s$ a mean value of $O(1/\alpha_s) \simeq$ 3 gluons) resulting in a
mean charged hadron multiplicity of about 20 
\item ``flavour democracy'' resulting in a significantly higher fraction of
strange hadrons in $I$-induced than in standard DIS-events
\item a higher total transverse momentum of $\simeq$5 GeV/$\eta$-unit as
compared to $\simeq 2$ GeV/$\eta$-unit for DIS-events \cite{15}
\end{itemize}

A typical MC-event is shown in fig.~3.


\begin{figure}[ht]
 \begin{minipage}[t]{7cm}
     \makebox[0cm][t]{}        
     \begin{center}
       \includegraphics[width=0.95\textwidth]{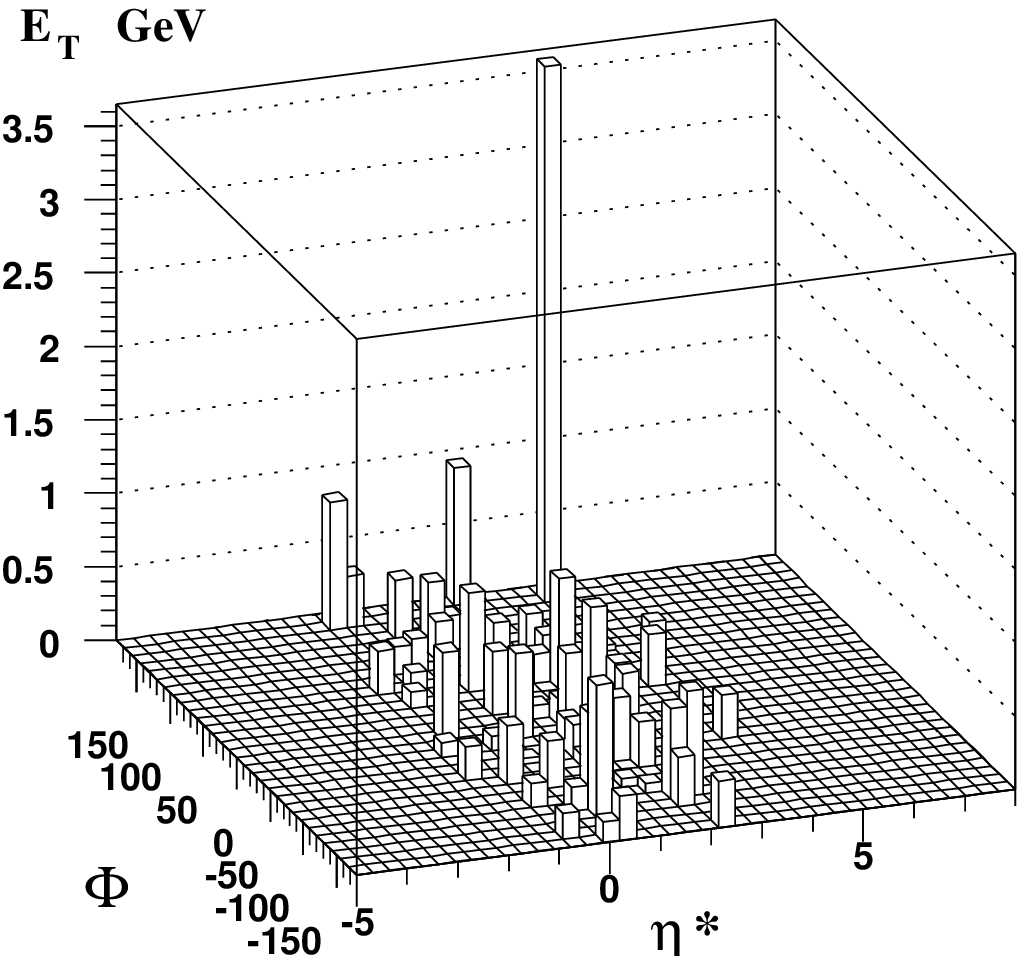}
     \end{center}
 \end{minipage}
 \hspace{0.5cm}
 \hfill
 \begin{minipage}[t]{7cm}
   \vspace{0.1cm} 
   \caption[]{\small Distribution of the transverse energy $E_T$ in 
              pseudorapidity $(\eta)$-azimuthal $(\phi)$-plane in the 
              hadronic CMS for a typical instanton-induced HERA-event 
              generated by QCDINS $(x = 0.0012, Q^2 = 66 \textrm{ GeV}^2, 
              p_T(\textrm{Jet}) = 3.6 \textrm{ GeV})$ after typical 
              detector cuts. Clearly recognizable are the current jet at 
              $\phi = 160^\textrm{{\small o}}, \eta \simeq 3$ and the
              instanton band at $0 \stackrel{<}{\sim} \eta
              \stackrel{<}{\sim} 2$. From Ref. \cite{11}} 
\label{abb01} 
\end{minipage}
\end{figure}

Search strategies for $I$-induced events in a background of normal DIS
events based on Monte Carlo data have been discussed by Carli, Gerigk,
Ringwald and Schrempp \cite{11}. The aim is to enrich $I$-induced events in
a data sample by cutting on selected observables while optimizing the
separation power, defined as the ratio of the detection efficiencies for
$I$-induced and DIS-events; in addition, a lower limit of 10$\%$ for the
$I$-detection efficiency is required. Fig.~4 shows the optimization for the
number of charged hadrons contained in the $I$-band after removal of the
current jet.

\begin{figure}[htb]
\begin{center}
\epsfig{file=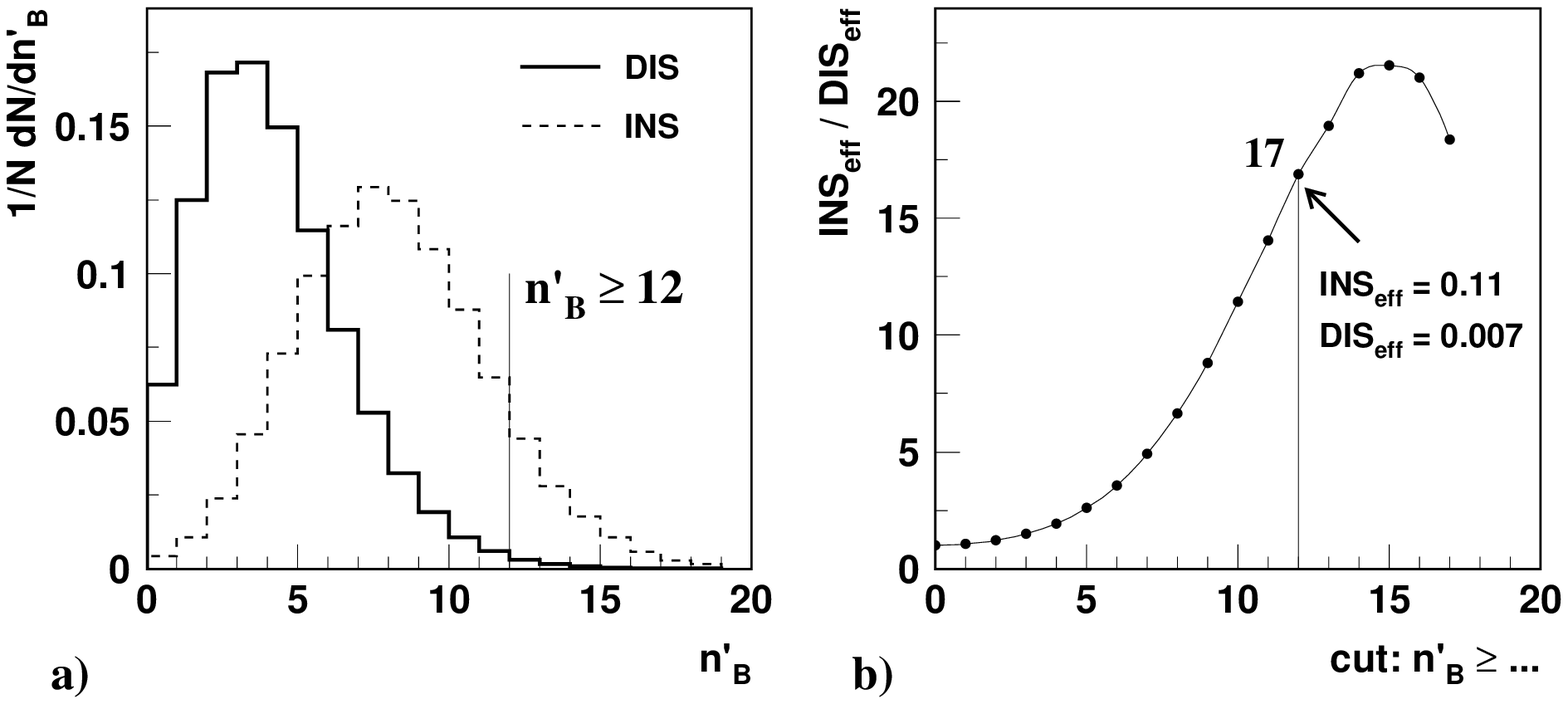,width=12cm}
\end{center}
\label{abb04}
\caption{\small a) Normalized MC-generated distributions of the number of
hadrons $n'_B$ in the instanton band for instanton induced (INS) and normal
events (DIS) after removal of the current jet hadrons. b) Maximum
separation power $INS_{eff}/DIS_{eff}$ as function of cuts on $n'_B$ (see
(a)) keeping the instanton efficiency $INS_{eff}$ at $\ge 10\%$. From
Ref. \cite{11}}
\end{figure} 

Cuts on single observables typically result in a separation power of
$\stackrel{<}{\sim} 20$ and, therefore, cuts on several
observables have to be combined in order to achieve a higher background
reduction. Correlations among the different variables, however, tend to
weaken the effect of combinations. For a set of 6 selected observables,
which are shown in fig.~5, a separation power of about 130 was obtained. 

\begin{figure}[h!]
\begin{center}
\epsfig{file=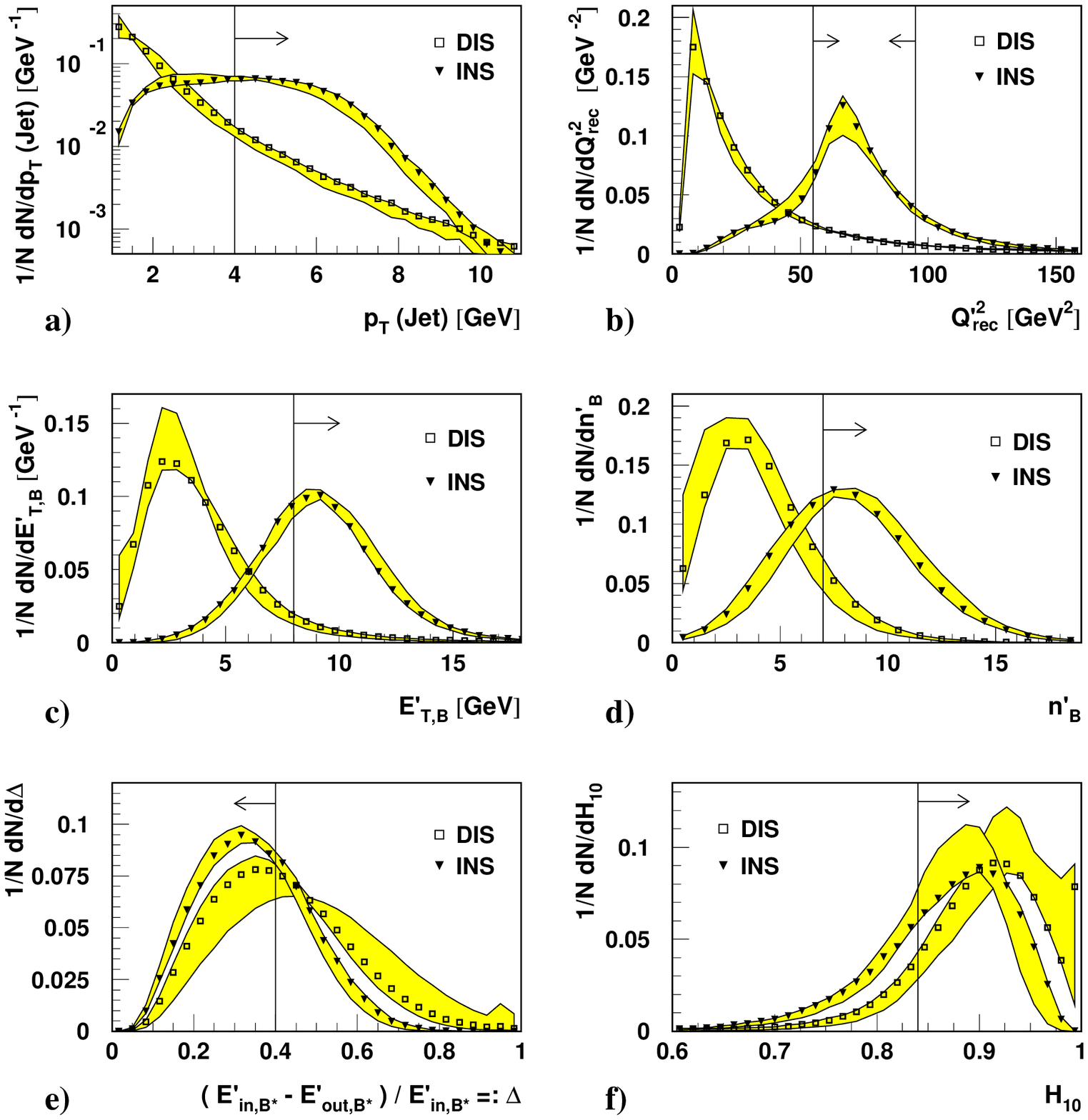,width=12cm}
\end{center}
\label{abb05}
\caption{\small Normalized distributions of characteristics observables
for normal DIS events (ARIADNE) and $I$-induced events (QCDINS +
HERWIG). Shaded bands indicate model uncertainties, arrows show optimized
cuts with the allowed region left of the cut. Variables: $p_T$ and $Q'^2$
refer to the current jet (compare fig.~1), $E'_{T,B}$ and $n'_B$ are the 
total transverse energy and number of hadrons, resp. in the $I$-band after
removal of the current jet, $\Delta$ is a measure of the isotropy of events
(see text). $H_{10}$ are the Fox-Wolfram-moments. From Ref. \cite{11}}
\end{figure}

\section{Recent H1 Results}

Based on this experience, the technique for enhancing $I$-induced
events has recently been applied by the H1 collaboration to real data
\cite{16}: Out of a set of 6 observables a subset of 3 observables has
been selected on which cuts are applied (called in the following ``primary
observables'') while the effect of these cuts on the distributions of the
other observables not subjected to cuts (called ``secondary observables'')
is studied. The preliminary results are based on a total integrated
luminosity of 15.8 pb$^{-1}$ collected by H1 in 1997 and corresponding to
about 260~k DIS-events within fixed cuts of $x > 10^{-3}, 0.1< y<0.6$
and $\Theta_e > 156^\textrm{{\small o}}$.

The primary observables chosen for the analysis were the number of hadrons
$n_b$ in the $I$-band, the sphericity $Sph$ of the event in the hadronic CMS
(both after removal of the current jet) and the squared momentum transfer
$Q'^2$ of the $I$-subprocess (c.f. fig.~1). The normalized distribution of
these observables are compared in fig.~6 with two quite different models
for DIS processes, i.e. the colour dipole model CDM \cite{17} and the
matrix element-parton shower-model MEPS \cite{18} as implemented in the Monte
Carlo ARIADNE resp. RAPGAP. Agreement on the level of 5-20$\%$ is
observed. The signal expected for $I$-induced events from QCDINS is $10^2 -
10^3$ below the DIS background.

A systematic investigation of the effect of cutting primary observables has
been carried out. The following combinations of cuts have been studied:
\mbox{$n_b > 5, 6, 7, 8, 9$}; $ Sph > 0.4, 0.5, 0.55, 0.6, 0.65;$ 
\mbox{$Q'^2 (\textrm{GeV}^2)$}: $95, 100, 105, 110,$ $115 < Q'^2 <
200$. The results have been classified in 3 scenarios: ``A'' for the
highest instanton efficiency $\epsilon_{INS}$, ``B'' for high instanton
efficiency at good separation power $\epsilon_{INS}/\epsilon_{DIS}$ and
``C'' for the highest separation power at an instanton efficiency $\ge$
$10\%$ (see table~1).


\begin{table}[h!]
\caption{\label{}}
\begin{center}
\begin{tabular}{|c|c|c|c|c|c|c|}
\hline
Scen & \multicolumn{3}{c|}{Cuts} & $\epsilon_{ins}$
& \multicolumn{2}{c|}{$\frac{\epsilon_{ins}}{\epsilon_{DIS}}$} \\
ario & $Q'^2(\textrm{GeV}^2)$ & Sph & $n_b$ & & CDM & MEPS \\
\hline
A & 95.0-200.0 & 0.40 & 5 & 32$\%$ & 35 & 34\\
\hline
B & 105.0-200.0 & 0.40 & 7 & 21$\%$ & 56 & 52\\
\hline
C & 105.0-200.0 & 0.50 & 8 & 11$\%$ & 86 & 71\\
\hline
\end{tabular}
\end{center}
\end{table}

\begin{figure}[h!]
\begin{center}
\epsfig{file=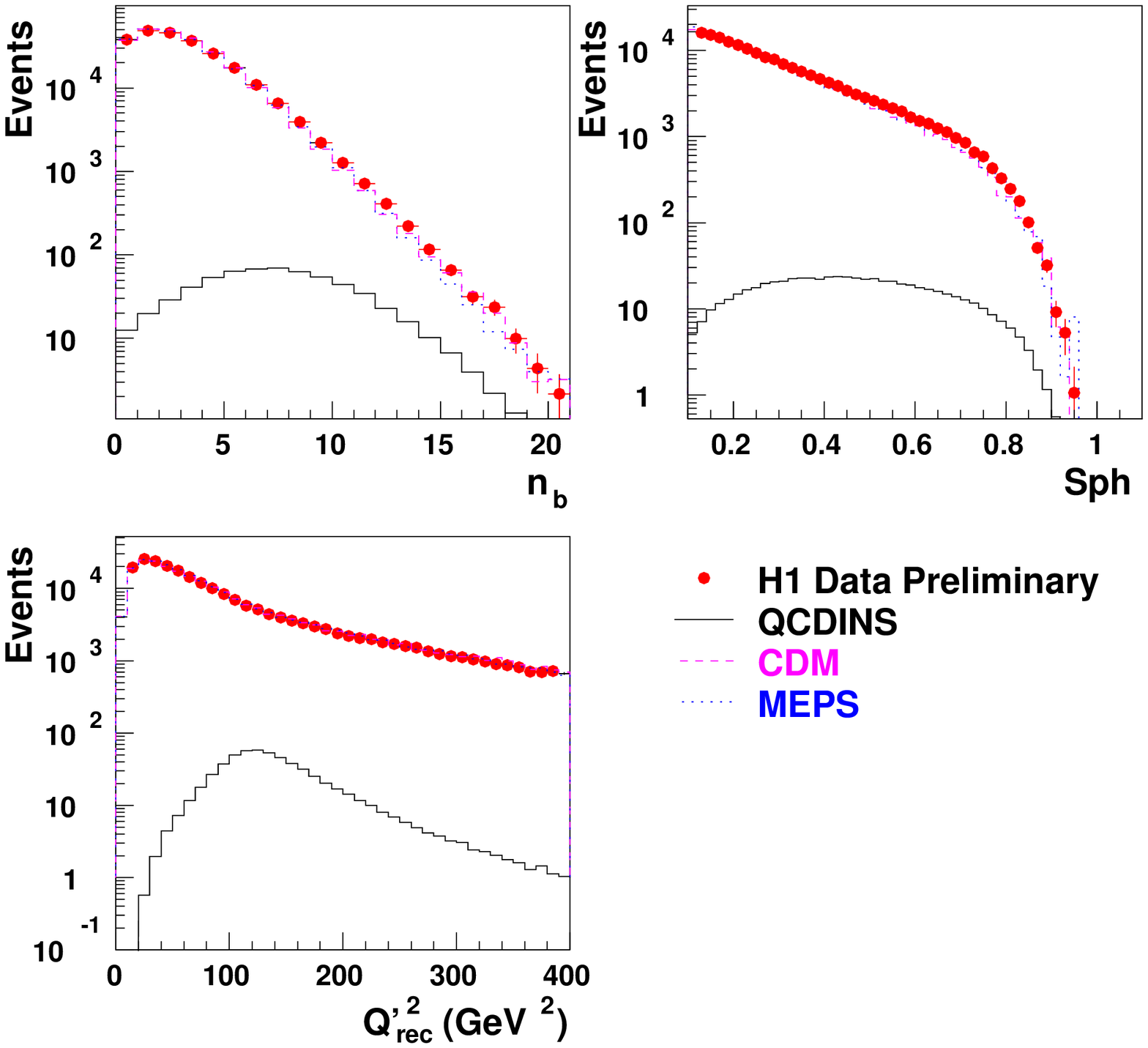,width=10cm}
\end{center}
\label{abb06}
\caption{\small Observables used to cut (``primary observables''):
Comparison of data with Monte Carlo (see text). From Ref. [16]}
\end{figure}

As example, the effect of cut C on primary observables is shown in
fig.~7. A background reduction by a factor of 600-800 is achieved. The
remaining 549 events are to be compared with 362 $\pm$ 25 events expected
from CDM and 435$\pm$30 events from MEPS. For the resulting distributions
the shape of the excess with respect to the DIS models is compatible with
the QCDINS expectation. It has to be noted, however, that the excess is of
the same magnitude as the difference between the two DIS models and that
also the QCDINS prediction is subject to systematic uncertainties
\cite{19}.

\begin{figure}[h!]
\begin{center}
\epsfig{file=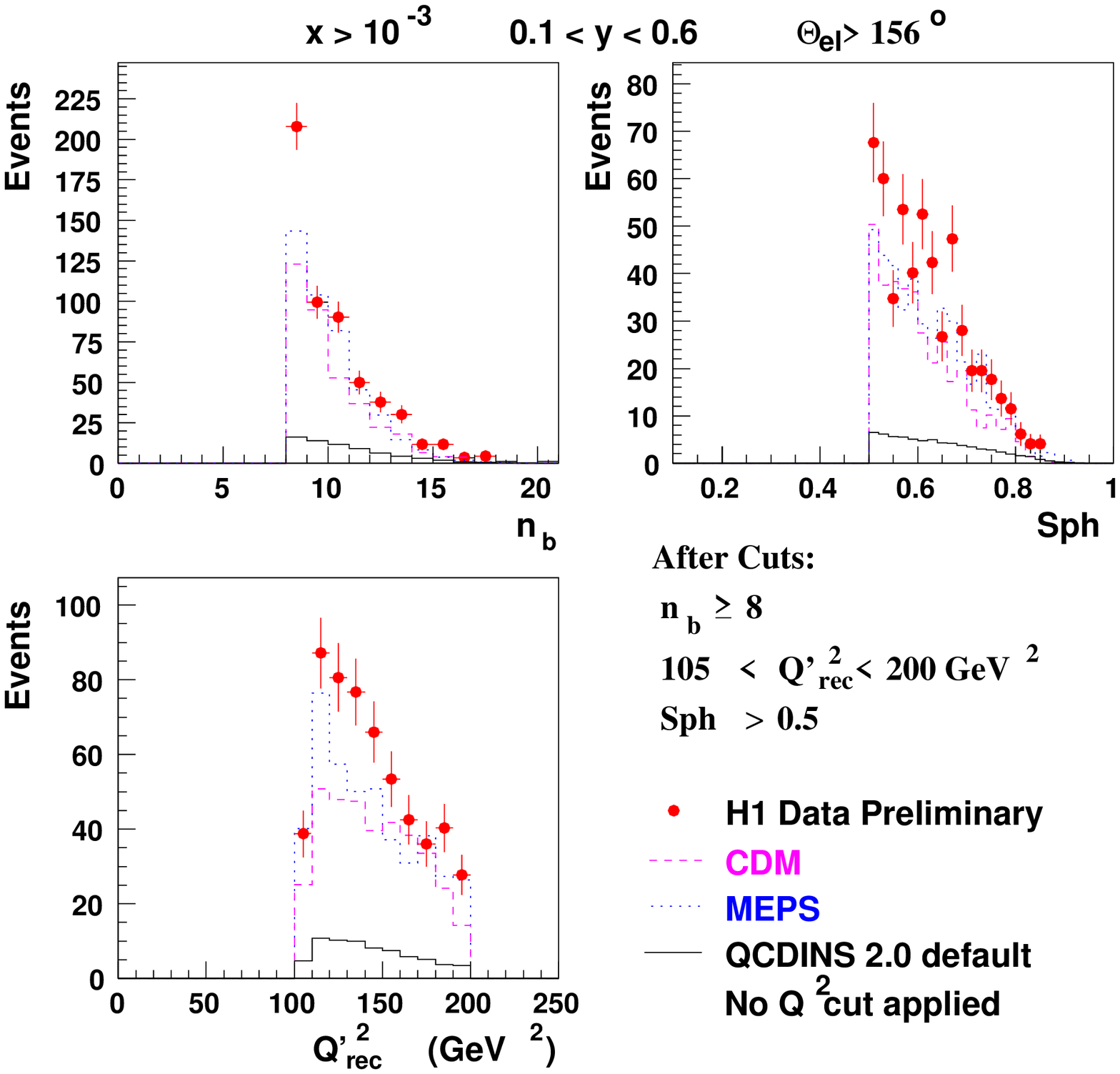,width=10cm}
\end{center}
\label{abb07}
\caption{\small Observables used to cut (``primary observables'') after
cuts. From Ref. [16]}
\end{figure}

As secondary observables were chosen the total transverse energy $E_{tb}$ of
hadrons contained in the $I$-band, the transverse energy $E_{t jet}$ of the
current jet and a third quantity $\Delta_b$, which is a measure for the
isotropy of events \cite{20,21}. It is obtained by finding the two axes with
respect to which the projection of all hadronic momenta (after removal of
the current jet) is minimal ($E_{in}$) resp. maximal ($E_{out}$) and
forming $(E_{in} - E_{out})/E_{in} = \Delta_b$. $\Delta_b$ measures the
$E_t$-weighted azimuthal isotropy; it takes small values for isotropic
events, expected for $I$-induced reactions, and large values for pencil-like
2-jet-events, expected to constitute a major fraction of the DIS-background.

The effect of cut C on the secondary observables is shown in fig.~8. While an
excess of events over both DIS Monte Carlos is observed also here,
differences in the shape of the distributions are recognizable; neither the
shape of the DIS Monte Carlos nor the shape of the signal expected from
QCDINS are well reproduced by the data.

\begin{figure}[h!]
\begin{center}
\epsfig{file=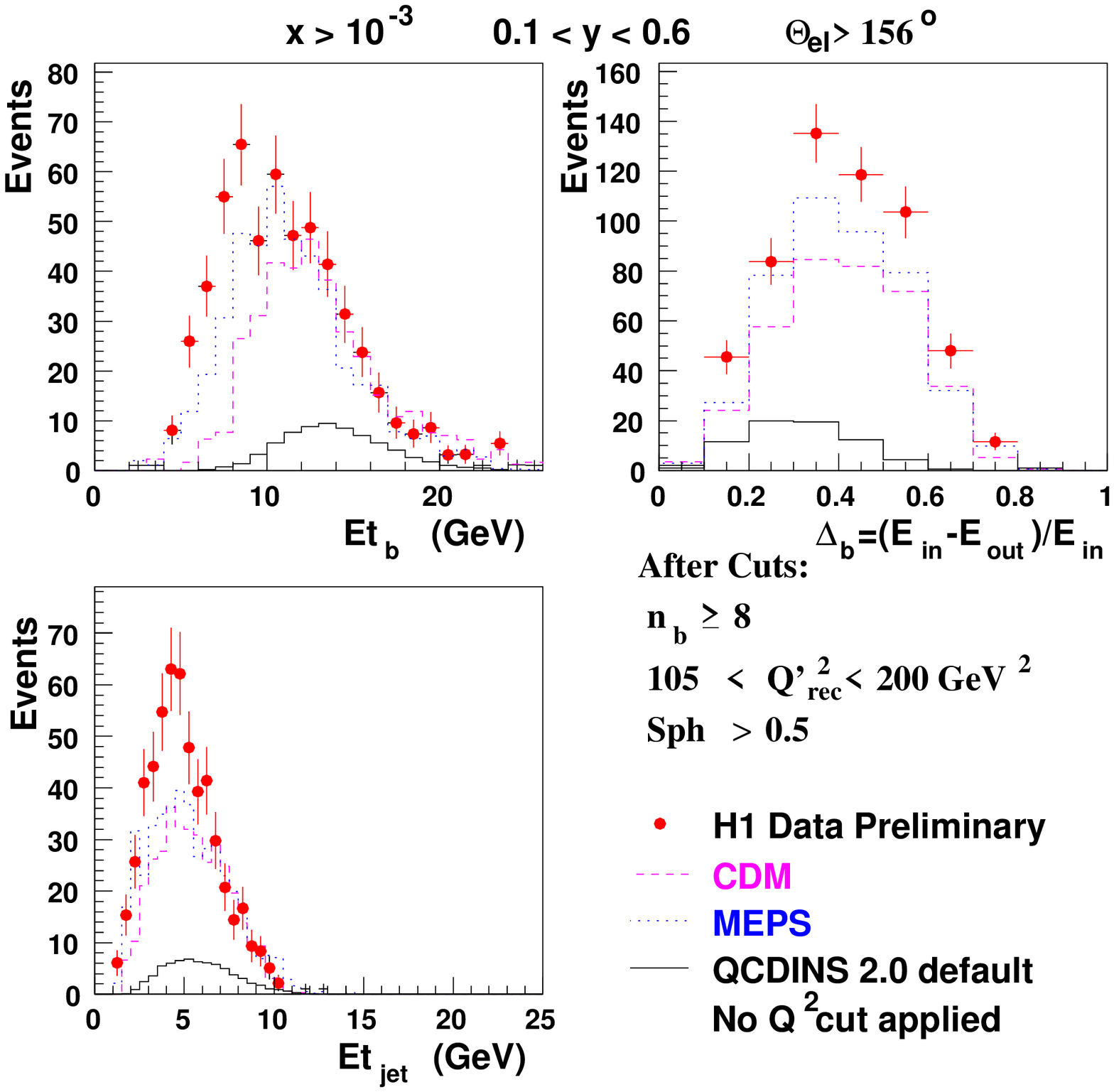,width=10cm}
\end{center}
\label{abb08}
\caption{\small Observables not used to cut (``secondary observables'')
after cuts on ``primary observables''. From Ref. \cite{16}}
\end{figure} 

These results do not yet allow any firm conclusion on a possible
contribution of $I$-induced processes to deep inelastic electron proton
scattering. Nevertheless, the observed excess of events and the gross
similarity between data and predictions are intriguing and motivate the
continuation of these studies with higher statistics and further refined
techniques. The ultimate signal of $I$-induced effects would consist, of
course, in demonstrating the violation of helicity-conservation in
DIS-events.

\section{Acknowledgements}

I would like to thank the Organizers of the Crimean Summer School Seminar,
in particular L. Jenkovszky, for an interesting and pleasant stay on the
Crimea. I am very grateful to G.~Grindhammer, A.~Ringwald and F.~Schrempp
for a critical reading of the manuscript and valuable comments.

\newpage

$^\ast$ Talk given at the Crimea Summer School Seminar ``New Trends in High
Energy Phyics'', May 27 - June 4, 2000, Yalta (Mishkor)


{\small
\begin{thebibliography}{99}
\bibitem{1}
A.~M. Polyakov, Phys.~Lett. 59B (1975) 82, 
A.~A. Belavin, A.~M. Polyakov, A.~S.~Schwartz, Yu.~S. Tyupkin,
Phys.~Lett. 59B (1975) 85
\bibitem{2}
G.~'t~Hooft, Phys.~Rev.~Lett. 37 (1976) 8, 
G.~'t~Hooft, Phys.~Rev. D14 (1976) 3432
\bibitem{3}
A.~Ringwald, Nucl.~Phys. B330 (1990) 1, 
O.~Espinosa, Nucl.~Phys. B343 (1990) 310
\bibitem{4}
S.~L.~Adler, Phys.~Rev. 177 (1969) 2426, 
J.~S.~Bell, R.~Jackiw, Nuov.~Cim.~60~(1969)~47
\bibitem{5}
T.V.~Sch\"afer, E.V.~Shuryak, Rev. Mod. Phys. 70 (1998) 323
\bibitem{6}
I.~Balitsky, V.~Braun, Phys.~Lett. B438 (1993) 237
\bibitem{7}
S.~Moch, A.~Ringwald, F.~Schrempp, Nucl. Phys. B507 (1997) 134
\bibitem{8}
A.~Ringwald, F.~Schrempp, Phys.~Lett. B438 (1998) 217
\bibitem{9}
A.~Ringwald, F.~Schrempp, Phys.~Lett. B459 (1999) 249
\bibitem{10}
A.~Ringwald, F.~Schrempp, \mbox{hep-ph/9411217}, in Quarks '94,
eds. D.~Yu.~Grigoriev et al., World Scientific, Singapore 1995
\bibitem{11}
T.~Carli, J.~Gerigk, A.~Ringwald, F.~Schrempp, DESY~00-067,
\mbox{MPI-PhE/99-02}, \mbox{hep-ph/9906441}, in Monte Carlo Generators for
HERA Physics, eds. A.~T.~Doyle et al.
\bibitem{12}
D.~A.~Smith, M.~J.~Teper, Phys.~Lett. D58 (1998) 014505
\bibitem{13}
S.~Capitani, M.~L\"uscher, R.~Sommer, H.~Wittig (ALPHA Coll.),
Nucl. Phys. B544 (1999) 669 
\bibitem{14}
A.~Ringwald, F.~Schrempp, hep-ph/9911516, Comput. Phys. Commun. (1999) in
print 
\bibitem{15}
H1 Collaboration, S.~Aid et al., Phys. Lett. B356 (1995) 118
\bibitem{16}
S.~Mikocki (H1 Collaboration), 8th International Workshop on Deep Inelastic
Scattering (DIS2000) in Liverpool, UK, June 2000, \mbox{hep-ex/0007008}
\bibitem{17}
B.~Andersson, G.~Gustafson, L.~L\"onnblad, U.~Petterson, Z. Phys. C43
(1989) 625
\bibitem{18}
M.~Bengtsson, T.~Sj\"ostrand, Z. Phys. C37 (1998) 465
\bibitem{19}
A.~Ringwald, F.~Schrempp, 8th International Workshop on Deep
Inelastic Scattering (DIS2000) in Liverpool, UK, June 2000, DESY~00-089,
hep-ph/0006215
\bibitem{20}
M.~Gibbs et al., in Proc. of the Workshop: Future Physics at HERA,
eds. G.~Ingelman et al., Vol. 1, p. 509, Hamburg 1996
\bibitem{21}
J.~Gerigk, Diploma Thesis, MPI-PhE/98-20
\end{thebibliography}}
\end{document}